\documentclass{tran-l}

\DeclareMathOperator*{\Var}{Var}
\DeclareMathOperator*{\Cov}{Cov}
\newcommand{\E}{\mathbb{E}}

\usepackage{graphicx}

\begin{document}

\title{Effective and simple VWAP options pricing model}

\author{ALEXANDER BURYAK}

\address{aburyak@bigpond.net.au}

\author{IVAN GUO}

\maketitle

\begin{abstract}
Volume weighted average price (VWAP) options are a popular security type in many countries, but despite their popularity very few pricing models have been developed so far for VWAP options. This can be explained by the fact that the VWAP pricing problem is set in an incomplete market since there is no underlying with which to hedge the volume risk, and hence there is no uniquely defined price. Any price, which is obtained will include a market price of volume risk which must be determined from the corresponding volume statistics. Our analysis strongly supports the hypothesis that the empirical volume statistics of ASX equities can be described reasonably well by fitted gamma distributions. Based on this observation we suggest a simple gamma process-based model that allows for the exact analytic pricing of VWAP options in a rather straightforward way.
\end{abstract}


\section{Introduction}

The volume weighted average price (VWAP) occurs frequently in finance. It is an average price which gives more weight to periods of high trading than to periods of low trading in its calculation. A broker's daily performance is frequently measured against the VWAP and it is becoming increasingly popular for institutional investors to place buy and sell orders at the VWAP. The VWAP also appears in Australian taxation law as part of
determining the prices of share buy-backs in publicly listed companies (Woellner {\it et al.} (2009)). 

Most of the existing literature on VWAP focuses on strategies and algorithms
to execute orders as close as possible to the VWAP price (see e.g. Konishi (2002), Bialkowski {\it et al.} (2008), Fuh {\it et al.} (2010), Frei \& Westray (2013)). 
On the other hand, surprisingly few results on actual pricing methodologies related to VWAP options have been published (Stace (2007), Novikov {\it et al.} (2014)). 
This can be explained by pointing out that the VWAP pricing problem is set in an incomplete market since there is no underlying with which to hedge the volume risk, and hence there is no uniquely defined price. Any price obtained will include a market price of volume risk which must be determined from the corresponding volume statistics.

In this paper, we propose a new model to price VWAP options in which the volume data is modelled by a gamma process. Exact closed-form expressions are derived for the first two moments of the VWAP, which may be used price VWAP options via well-known moment matching techniques. We then compare our results against the technique suggested by Stace (2007) 
as well as with Monte Carlo modelling results.

The rest of this paper is organised as follows. Section 2 briefly describes some of the previously suggested models for volume data.
Section 3 justifies our choice of the gamma process as the preferred volume model by presenting goodness-of-fit results and other analyses of volume data. Section 4 formally introduces our model for both stock price (lognormal) and stock volume (gamma).
Section 5 presents the main results of the paper, which include closed-form expressions for the VWAP moments and option prices (based on a moment matching technique) as well as a comparison to Monte Carlo results.
Section 6 contains some concluding remarks.
Detailed derivations of VWAP moments, for both discrete and continuous-time cases, can be found in Appendix A.

\section{Previously suggested models for volume process.}

It is common to model the underlying stock price $S_t$ using the standard geometric Brownian motion. On the other hand, the choice of model for the volume traded $V_t$ is far less obvious. A few versions of the volume process $V_t$ have been suggested in the literature. We shall briefly outline some of these existing approaches before selecting our own.

For example, Stace (2007) 
has considered the following mean reverting volume process:
\begin{gather} \label{Stace_V_evolution}
{d V_t = \lambda(V_{\rm{mean}} - V_t)dt + \beta V_t^f dW,}
\end{gather}
where $V_{t=0}$ is given, $\lambda$ is the speed of mean reversion, $V_{\rm{mean}}$ is the long term average of the
volume process, $\beta$ is the volatility of the volume process, $W$ is a standard Brownian motion (which may be partially correlated to the stock price process Brownian motion), and $f$ is either $1$ or $0.5$.

In the more recent work of Novikov {\it et al.} (2014) 
a different class of volume processes has been suggested:
\begin{gather} \label{Novikov_V_evolution}
V_t = X_t^2 + \delta,\qquad
d X_t = \lambda(X_{\rm{mean}} - X_t) dt + \beta dW,
\end{gather}
where $X$ is a standard Ornstein-Uhlenbeck (OU) process with $\lambda$, $X_{\rm{mean}}$ and $\beta$ being its speed of mean-reversion, level of mean-reversion and OU volatility, respectively.

Yet another $V_t$ model, also based on presence of a second Brownian motion in underlying dynamics was suggested in Fuh {\it et al.} (2010). 
However, all these and many other VWAP-related publications are predominantly concentrated on the description of their pricing or trading algorithms, with little attention given to any justification of the corresponding volume process model choice or any comprehensive empirical analysis of volume statistics. A notable exception is Frei \& Westray (2013), 
in which a considerable amount of attention is paid to such a justification, as well as to outlining of useful approaches for empirical checks. Importantly, Frei \& Westray (2013) 
argues that there is substantial empirical evidence to suggest that $V_t$ can be modelled by i.i.d.\! gamma random noises. This choice may look unusual for a financial model, but as Brody {\it et al.} (2008) 
demonstrates, gamma processes actually have a broad range of applications in many areas of insurance and finance, where cumulative processes are involved. These include the modelling of aggregative claims, credit portfolio losses, defined benefit pension schemes and so on.

\section{Empirical analysis of volume data.}

In this section we present our empirical analysis justifying the use of gamma variables to model underlying dynamics of traded volume. This, in turn, allows us to progress with the development of a VWAP option pricing model in the next section.
In particular, we analyse the price and the volume data series for a few ASX stocks.

Data for our analysis is obtained from Bloomberg and covers the period from 15/02/2013 until 28/08/2013. Each original data point represents a traded volume and the corresponding VWAP price for a 10 minute interval. Typically 38 traded volume data points $V_i$ per day are available (6 per hour for 6 hours, plus one pre-market trading [10 am] point and one post-market [4.10 pm] point). Thus, on average, we have about 5130 data points per stock. This number varies slightly from stock to stock, because occasionally some data points are missing (e.g., due to the lack of trading within certain 10 min intervals).

For each equity volume data set we construct 4 secondary sets $V_i^{(L)}$ by combining volumes of $L$ consecutive points (and thus amalgamating the original $L$-point groups to single points in the newly derived data sets), where $L = 5, 10, 20$ and $40$, corresponding to approximately $1/8$, $1/4$, $1/2$ and $1$ day incremental volumes, respectively. Then gamma distributions are fitted to these newly constructed amalgamated volume data sets. We recall that the standard gamma distribution $\Gamma(\alpha,\theta)$ has a mean $\alpha \theta$ and variance $\alpha \theta^2$ (see e.g., Brody {\it et al.} (2008) 
for a detailed discussion of gamma distribution properties and additional relevant references).

Due to properties of the gamma distribution, if the original (non-amalgamated) volume data follows a true gamma distribution, then its $\theta$ parameter would stay constant with respect to $L$, its $\alpha$ parameter would scale so that $\alpha(L)/L$ is constant, and all time series and autocorrelation coefficients would satisfy $C_{\rm auto}(L) \ll 1$. In addition to the analysis of $\theta(L)$, $\alpha(L)$ and $C_{\rm auto}(L)$, we perform two goodness-of-fit tests: Anderson-Darling (A-D) and Kolmogorov-Smirnov (K-S), based on the hypothesis of $V_i^{(L)}$ having gamma distribution. The A-D and K-S goodness-of-fit analysis was conducted using the Mathematica 9.0 software package and below we only report the corresponding $P$-values (with higher values meaning higher probability of the hypothesis being correct).

\begin{table}[ht]
  \caption{Statistical analysis of CBA, WDC and FMG stock volume data (data source: Bloomberg) and its comparison with analysis of synthetic (computer-generated) data sample (with gamma distribution parameters $\alpha = 1.0$ and $\theta = 0.2 \times 10^6$). For stock data each original data point represents a traded volume and the corresponding VWAP price for a 10 minute interval (approx. 5130 points are available for each stock; for synthetic data we also generate 5130 points). Then 4 secondary sets $V_i^{(L)}$ are constructed by combining volumes of $L$ consecutive points ($L = 5, 10, 20, 40$). Then gamma distributions are fitted to these amalgamated volume data sets (i.e., we obtain fitted gamma distribution parameter values $\theta(L)$ and $\alpha(L)$ for each set) and autocorrelation coefficients $C_{\rm auto}(L)$ are also calculated. In addition, we perform two goodness-of-fit tests: Anderson-Darling (A-D) and Kolmogorov-Smirnov (K-S), based on the hypothesis of $V_i^{(L)}$ having a gamma distribution and report the corresponding $P$-values (with higher values meaning higher probability of the hypothesis being correct).}
{\begin{tabular}{|c|c|c|c|c|c|c|c|} \hline
& & & & & & & \\
${\rm stock/}$ & $L$ & $\theta(L)/10^6$ & $\alpha(L)$ & $\alpha/L$  & $C_{\rm auto}(L)$ & $P^{\rm(A-D)}$ & $P^{\rm(K-S)}$ \\
${\rm synthetic}$ & & & & & & &\\
& & & & & & &\\
\hline
& & & & & & &\\
${\rm CBA}$ & 5  & 0.33 & 3.23 & 0.65 & 39\% & 0.01\% & 0.11\% \\
${\rm CBA}$ & 10 & 0.54 & 3.95 & 0.39 & 21\% & 5.0\%  & 4.6\%  \\
${\rm CBA}$ & 20 & 0.73 & 5.84 & 0.29 & 8\% & 59.1\% & 49.4\%  \\
${\rm CBA}$ & 40 & 0.77 & 11.07 & 0.28 & 45\% & 90.8\% & 74.8\% \\
& & & & & & & \\
\hline
& & & & & & &\\
${\rm WDC}$ & 5  & 0.95 & 2.19 & 0.44 & 21\% & 0.00\% & 0.00\% \\
${\rm WDC}$ & 10 & 1.46 & 2.87 & 0.29 & 2\%  & 2.8\%  & 1.9\%  \\
${\rm WDC}$ & 20 & 1.77 & 4.73 & 0.24 & -14\%& 54.6\% & 31.6\% \\
${\rm WDC}$ & 40 & 1.55 & 10.82 & 0.27 & 48\%& 99.7\% & 98.7\% \\
& & & & & & & \\
\hline
& & & & & & &\\
${\rm FMG}$ & 5  & 2.58 & 2.82 & 0.56 & 40\% & 1.7\%  & 42.3\% \\
${\rm FMG}$ & 10 & 3.88 & 3.75 & 0.38 & 25\% & 37.4\% & 32.3\% \\
${\rm FMG}$ & 20 & 5.14 & 5.66 & 0.28 & 10\% & 73.1\% & 58.4\% \\
${\rm FMG}$ & 40 & 5.29 & 10.99 & 0.27 & 33\%& 93.9\% & 95.9\% \\
& & & & & & & \\
\hline
& & & & & & &\\
${\rm synthetic}$ & 5  & 0.20 & 4.99 & 1.00 &  2\% & 93.2\% & 90.4\% \\
${\rm synthetic}$ & 10 & 0.21 & 9.50 & 0.95 &  0\% & 99.9\% & 94.0\% \\
${\rm synthetic}$ & 20 & 0.22 & 17.65 & 0.88 & 2\% & 90.7\% & 89.6\% \\
${\rm synthetic}$ & 40 & 0.20 & 38.18 & 0.98 & 4\% & 84.2\% & 76.2\% \\
& & & & & & & \\
\hline
\end{tabular}}
\label{CBA}
\end{table}

Our results for CBA, WDC and FMG stocks are presented in Table 1 (all these stocks are part of the ASX 200 index). These results typically display significant deviations from true gamma distribution behaviours for smaller bucket size $L$, but strongly support the hypothesis of $V_i^{(L)}$ having gamma distributions for $L \ge 20$ (i.e., with half a day or longer averaging).

For the sake of giving the reader a better feel of how close our stock volume statistics hypothesis is to reality, we use the algorithm of Press {\it et al.} (2007) 
to generate the same number of sample points of ``true'' gamma-distributed noise (5130 points) and repeat our analysis. The corresponding results are also included in Table 1.

\begin{figure}[h]
\begin{center}
\resizebox*{8cm}{!}{\includegraphics{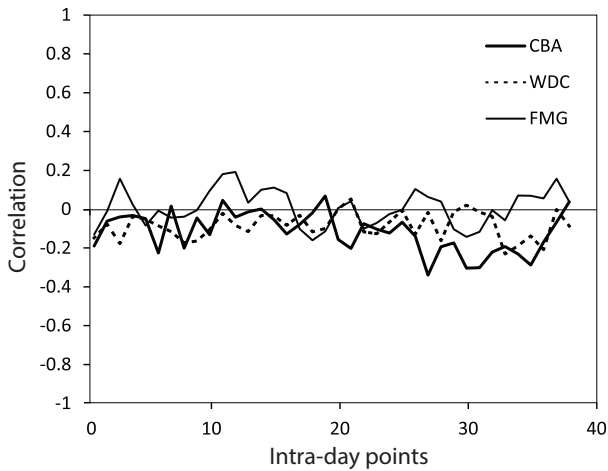}}%
\caption{Correlation between cumulative volume $V^{(c)}_i = \sum_{j = 1}^i V_j$ and the corresponding relative trading volume $V_i$ for CBA, WDC and FMG. Each correlation point is calculated for data sets obtained for the same time period $[t_i,t_i+\Delta t]$ of available trading days.}%
\label{Fig1}
\end{center}
\end{figure}

Finally we note that some other tests suggested by Frei \& Westray (2013) 
are also performed. The most important one being a test demonstrating a low correlation between cumulative ($\sum_{j = 1}^i V_j$) and incremental volumes ($V_i$) within the same time period of a trading day $[t_i,t_i+\Delta t]$ (i.e. day-to-day independence of $\sum_{j = 1}^{i} V_j$ and $V_i$; see Frei \& Westray (2013) for details). 
Figure 1 shows the corresponding intra-day correlations for CBA, WDC and FMG stocks, which are indeed reasonably low.

All results reported in this section strongly support the hypothesis that at least for the averaging interval of half a day or longer the stock volume dynamics can indeed be described as the standard gamma process: if the volume traded within the time period $[t_i,t_i+\Delta t]$ is given by $V_{i}$, then we assume that it has the gamma distribution $\Gamma(\alpha,\theta)$, with mean $\alpha \theta$ and variance $\alpha \theta^2$, where $\alpha$ depends linearly on the averaging period $\Delta t$. Due to the independent increment property of the gamma process, volumes traded in disjoint time intervals of equal length are i.i.d.\! gamma variables.

\section{Model for VWAP option pricing.}

We now formally define a new model for the VWAP option pricing using a gamma process for volume dynamics.
We work under the filtered probability space $(\Omega, \mathcal{F}, \mathbb{F}, \mathbb{Q})$. The filtration $\mathbb{F}=\{\mathcal{F}_t\}$ represents the flow of information available to market participants. In particular, it is the augmented filtration generated by a standard Brownian motion $W_t$ and a sequence of i.i.d.\! gamma variables $V_1,\ldots, V_N \sim \Gamma(\alpha,\theta)$. For every $i=1,\ldots,N$, the gamma variable $V_i$ is $\mathcal{F}_{t_i}$-measurable where $t_i:=i\Delta t$ for some fixed time increment $\Delta t$.
Note that the process $W_t$ and the random variables $\{V_i\}$ are assumed to be independent. For convenience we choose $W_t$ to be the Brownian motion under the risk-neutral pricing measure. So the measure $\mathbb{Q}$ is the product of the risk-neutral pricing measure and the real-world measure associated with $\{V_i\}$.

To summarise: the \emph{stock price} process $S_t$ is given by the standard geometric Brownian motion
\begin{gather} \label{S_evolution}
d S_t = r S_t dt + \sigma S_t dW_t,
\end{gather}
where $r$ is the risk-free interest rate and $\sigma$ is the volatility. Note that the discounted stock price $e^{-rt}S_t$ is a martingale under $\mathbb{Q}$.

The \emph{volumes traded} $V_{t_i}$ traded during the time periods $[t_{i-1},t_{i}]$ (where $t_{i} = i\Delta t$) are directly modelled by the i.i.d.\! gamma variables,
\begin{gather} \label{Z_evolution}
V_{t_i} := V_i \sim \Gamma(\alpha,\theta),
\end{gather}
with mean $\alpha \theta$ and variance $\alpha \theta^2$.

Now we can define the VWAP on a time interval $[0, T = N \Delta t]$ as
\begin{align} \label{VWAPd}
S^{\rm (VWAP)} = \frac{\sum^N_{i=1} S_i V_i}{\sum^N_{i=1} V_i}. 
\end{align}
This paper will focus on the discrete-time formulation of VWAP \eqref{VWAPd} which is a weighted average of $N$ incremental trade volumes as opposed to a weighted integral over time. For completeness, the results for the continuous-time limit can be found in Appendix A.
Note that we will consider the standard call and put VWAP options:
\begin{align} \label{VWAP_payouts}
C &= \max(S^{\rm (VWAP)} - K,0),\\
P &= \max(K - S^{\rm (VWAP)}, 0).
\end{align}


\section{Closed-form pricing formulas for VWAP options}

It is natural to start further analysis with a brief description of currently known approaches (Stace (2007), Novikov \& Kordzakhia (2013), Novikov {\it et al.} (2014)). 
for VWAP option pricing. Note that two of these works (Stace (2007), Novikov {\it et al.} (2014)) 
utilise a well-known moment-matching technique, which works reasonably well for problems such as pricing Asian options (see e.g., Hull (2006)). 
 In this technique, we match the first two moments of the VWAP to the corresponding moments of the standard lognormal process (for which all pricing formulas are well-known) which have the following form (see, e.g., Glasserman (2003)):
\begin{align} \label{Lognormal_moments}
M_1^{(LN)}&={\E}[S_t] = S_0 \exp{(r t}),\\
M_2^{(LN)}&={\Var}[S_t] = S_0^2 [\exp{(\sigma^2 t)} - 1] \exp{(2 r t)},
\end{align}
where all notations are the same as (\ref{S_evolution}). After matching calculated VWAP moments to the lognormal moments (i.e. taking $M_1^{(LN)}=M_1^{(VWAP)}$ and $M_2^{(LN)}=M_2^{(VWAP)}$), one can use the standard Black formulas for options pricing with the forward $F_0 = M_1$ and the volatility $\sigma^2 =\ln{(M_2/M_1^2+1)}/T$ (see, e.g., Hull (2006)). 
The approach is known to work very well for low volatilities ($\sigma \le 0.2$) and reasonably well for higher volatility levels ($0.2 < \sigma \le 0.4$).

In the work of Stace (2007) 
the standard moment matching approach (described above) is supplemented by further approximations during the calculation of the VWAP moments, utilising the following expressions:
\begin{align} \label{Ratios}
{\rm \E}\left(\frac{Y}{Z}\right) &\approx \frac{{\rm \E}\left(Y \right)}{{\rm \E}\left(Z \right)}
- \frac{{\rm Cov}\left(Y, Z \right)}{\left[{\rm \E}\left(Z \right)\right]^2} + \frac{{\rm \E}\left(Y \right)}{\left[{\rm \E}\left(Z \right) \right]^3} {\rm Var}(Z),\\
{\rm Var}\left(\frac{Y}{Z}\right) &\approx \left(\frac{{\rm \E}\left(Y \right)}{{\rm \E}\left(Z \right)}\right)^2 \left( \frac{{\rm Var}\left(Y \right)}{\left[{\rm \E}\left(Y \right)\right]^2}
 +\frac{{\rm Var}\left(Z \right)}{\left[{\rm \E}\left(Z \right)\right]^2} - 2 \frac{{\rm Cov}\left(Y,Z \right)}{{\rm \E}\left(Y \right){\rm \E}\left(Z \right)} \right),\label{Ratios2}
\end{align}
where all terms may be computed explicitly after lengthy, but rather straightforward calculations. Note that these approximations are rather well-known (see e.g., Mood {\it et al.} (1974), p.\! 181). 

Another possible approach for VWAP option pricing was recently outlined in Novikov \& Kordzakhia (2013) 
and is essentially based on an upper/lower bounds approach, which is also well-known for Asian options (see, e.g., Curran (1994), Rogers \& Shi (1995), Thompson (2000), Lord (2006) to name a few). 
Although more accurate than the moment-matching approach for higher volatility values ($\sigma > 0.2$), this approach typically does not allow closed-form analytical representations (some numerical search for minimum/maximum values over one or more parameters is present).

At this point we choose to follow the simpler and more tractable moment matching method of pricing, aiming to get a simple closed-form analytical result at the end. Thus we need to calculate the first and the second moments for $S^{\rm (VWAP)}$ which is given by Eq. (\ref{VWAPd}), where evolution of $S_t$ and $V_t$ are defined by Eq. (\ref{S_evolution}) and Eq. (\ref{Z_evolution}), respectively.

It is important to note that, in contrast to lognormal volume dynamics models, the system (\ref{S_evolution})--(\ref{Z_evolution}) allows us to obtain a closed-form moment matching-based pricing solution, both with and without using an additional approximation (\ref{Ratios})--(\ref{Ratios2}).
Detailed derivation of the solution to the system (\ref{S_evolution})--(\ref{Z_evolution}) can be found in Appendix A. Here we only present some final expressions for the volatility correction factor $R^{\rm (VWAP)} \equiv \sigma_D^{\rm (VWAP)}/\sigma_D^{\rm (AA)}$, (where $\sigma_D^{\rm (AA)}$ represents the standard moment matching-based implied volatility result for arithmetic averaging Asian options) noting that compact closed-form asymptotic expansion formulas are available in the limit $T/N \ll 1$:
\begin{align} \label{Exact}
R^{\rm (VWAP)}_{\rm Exact} &= \sqrt{\frac{N (3 + \alpha + 2 \alpha N)}{(1+2 N) (1 + \alpha N)}+ O(T/N)},\\
\label{Stace_like}
R^{\rm (VWAP)}_{\rm Stace} &= \sqrt{\frac{(3 + \alpha + 2 \alpha N)}{(\alpha + 2 \alpha N)}+ O(T/N)}.
\end{align}

Also note that for our gamma-process model $F^{\rm(VWAP)} = F^{\rm (AA)}$, i.e., no VWAP forward correction is required in comparison to the corresponding arithmetic averaging Asian result.

 A few observations can be made from the analysis of (\ref{Exact})--(\ref{Stace_like}):\\
(i) the parameter $\theta$ is not present in the result (due to its absence in the relative variance formula for the gamma distribution: $\Var(\Gamma)/\E(\Gamma)^2 = 1/\alpha$),\\
(ii) in case of daily ($T/N = 1/252$) or more frequent averaging, all the extra terms $O(T/N)$ are negligible in comparison to the leading order terms for any value of $T$,\\
(iii) for up to $T/N \sim 0.5$, the dependence of $R^{\rm VWAP}$ on $T$ is weak (in comparison with its dependence on $\alpha$ or $N$) and may be ignored, at least as a first approximation.

We may now check the quality of the results (\ref{Exact})--(\ref{Stace_like}) by comparing them to direct Monte-Carlo (MC) simulations. The results are presented in Table 2.

\begin{table}[h]
  \caption{Comparison of the gamma model effective volatility results with the corresponding MC results for the following parameter values: $\sigma = 0.2$, $r = 0.05$, $\theta = 0.00067$, MC number of paths $= 10000000$, $T = 2/52$ (with $N=10$ averaging points). We note that the $R^{\rm (VWAP)}_{\rm Exact}$ column gives a ratio $\sigma^{\rm (VWAP)}_{\rm ExactD}/\sigma_D^{\rm (AA)}$ result based on the discrete-time exact formulas of Appendix A; the $R^{\rm (VWAP)}_{\rm Stace}$ column gives $\sigma^{\rm (VWAP)}_{\rm StaceD}/\sigma_D^{\rm (AA)}$ result based on the discrete-time ``Stace-like'' formulas of Appendix A; $R^{\rm (VWAP)}_{\rm MC}$ column gives the corresponding result of direct MC modelling $\sigma_{\rm MC}^{\rm (VWAP)}/\sigma_{\rm MC}^{\rm (AA)}$; and the last column provides an error estimate for the ``Stace-like'' ratio result $(R^{\rm (VWAP)}_{\rm Stace}-1)/(R^{\rm (VWAP)}_{\rm MC}-1)$.}
{\begin{tabular}{|c|c|c|c|c|} \hline
& & & & \\
$1/\alpha$ & $R^{\rm (VWAP)}_{\rm Exact}$ & $R^{\rm (VWAP)}_{\rm Stace}$ & $R^{\rm (VWAP)}_{\rm MC}$  & $\frac{(R^{\rm (VWAP)}_{\rm Stace}-1)}{(R^{\rm (VWAP)}_{\rm MC}-1)}$  \\
& & & & \\
\hline
& & & & \\
0.00 & 1.0000 & 1.0000 & 1.0000 $\pm$ 0.0002 & -      \\
0.02 & 1.0004 & 1.0014 & 1.0004 $\pm$ 0.0002 & 3.40   \\
0.20 & 1.0042 & 1.0142 & 1.0042 $\pm$ 0.0002 & 3.36   \\
0.50 & 1.0102 & 1.0351 & 1.0102 $\pm$ 0.0002 & 3.43   \\
0.75 & 1.0148 & 1.0351 & 1.0150 $\pm$ 0.0002 & 3.49   \\
1.00 & 1.0193 & 1.0522 & 1.0194 $\pm$ 0.0002 & 3.56   \\
1.20 & 1.0227 & 1.0823 & 1.0228 $\pm$ 0.0002 & 3.62   \\
1.50 & 1.0276 & 1.1019 & 1.0276 $\pm$ 0.0002 & 3.69   \\
1.80 & 1.0322 & 1.1212 & 1.0322 $\pm$ 0.0002 & 3.76   \\
2.00 & 1.0351 & 1.1339 & 1.0352 $\pm$ 0.0002 & 3.80   \\

& & & & \\
\hline
\end{tabular}}
\label{Gamma}
\end{table}

The parameter values in Table 2 are as follows: $\sigma = 0.2$, $r = 0.05$, $\theta = 0.00067$, MC number of paths $= 10000000$, $T = 2/52$ (with $N=10$ averaging points, i.e. a two-week term with daily averaging). More specifically:
the $R^{\rm (VWAP)}_{\rm Exact}$ column gives a ratio $\sigma^{\rm (VWAP)}_{\rm ExactD}/\sigma_D^{\rm (AA)}$ result based on the discrete-time exact formulas of Appendix A; the $R^{\rm (VWAP)}_{\rm Stace}$ column gives $\sigma^{\rm (VWAP)}_{\rm StaceD}/\sigma_D^{\rm (AA)}$ result based on the discrete-time ``Stace-like'' formulas of Appendix A; $R^{\rm (VWAP)}_{\rm MC}$ column gives the corresponding result of direct MC modelling $\sigma_{\rm MC}^{\rm (VWAP)}/\sigma_{\rm MC}^{\rm (AA)}$; and the last column provides an error estimate for the ``Stace-like'' ratio result $(R^{\rm (VWAP)}_{\rm Stace}-1)/(R^{\rm (VWAP)}_{\rm MC}-1)$.
Note that we omit a similar comparison for $R^{\rm (VWAP)}_{\rm Exact}$ ratio as it perfectly matches the MC result (i.e. is always within MC error margin).

We see a substantial disagreement of Stace approximation (\ref{Stace_like}) with numerics, whereas the exact solution (\ref{Exact}) matches MC results perfectly (within MC error bounds).

Table 3 provides the comparison of VWAP implied volatilities (based on the gamma model considered in this Section) and the corresponding vanilla option prices with their arithmetic Asian option counterparts. For all examples, the parameters are $\sigma = 0.2$,
$r = 0.05$ and $S = K = \$ 100.00$.

\begin{table}[h]
  \caption{Comparison of VWAP implied volatilities $\sigma^{\rm (VWAP)}_D$ and VWAP option prices $P^{\rm (VWAP)}_D$ with the corresponding arithmetic Asian volatilities $\sigma^{\rm (AA)}_D$ and option prices $P^{\rm (AA)}_D$ for some typical values of gamma parameter $\alpha$, option tenor $T$ and number of averaging points $N$. For all examples, the parameters are $\sigma = 0.2$,
$r = 0.05$ and $S = K = \$ 100.00$. }
{\begin{tabular}{|c|c|c|c|c|c|c|c|c|} \hline
& & & & & & & & \\
Type & $T$ & $\alpha$ & $N$  & $\sigma^{\rm (AA)}_D$ & $\sigma^{\rm (VWAP)}_D$
& $P^{\rm (AA)}_D$ & $P^{\rm (VWAP)}_D$ & $\frac{P^{\rm (VWAP)}_D}{ P^{\rm (AA)}_D} - 1$ \\
 & [years] &  &  & [$\%$] & [$\%$] & [$\$ $] & [$\$ $] & [$\%$] \\
& & & & & & & & \\
\hline
& & & & & & & & \\
put & 0.317 & 10 & 80 & 11.68 & 11.69 & 2.217 & 2.217 & 0.04   \\
call & 0.317 & 10 & 80 & 11.68 & 11.69 & 3.012 & 3.012 & 0.03   \\
put & 0.079 & 10 & 20 & 11.99 & 12.00 & 1.242 & 1.242 & 0.12   \\
call & 0.079 & 10 & 20 & 11.99 & 12.00 & 1.450 & 1.450 & 0.11   \\
put & 0.020 & 10 & 5 & 13.27 & 13.32 & 0.716 & 0.718 & 0.37   \\
call & 0.020 & 10 & 5 & 13.27 & 13.32 & 0.775 & 0.778 & 0.34   \\
put & 0.020 & 5 & 5 & 13.27 & 13.36 & 0.716 & 0.721 & 0.73   \\
& & & & & & & & \\
\hline
\end{tabular}}
\label{Examples}
\end{table}

The last three columns of Table 3 report arithmetic Asian and VWAP option prices and their relative difference for a typical for ASX stocks ($\alpha = 10.0$) and lowest observed ($\alpha = 5.0$) values of the parameter $\alpha$. Clearly this relative difference is quite noticeable for some examples.

\section{Conclusions and discussion}

In conclusion, we suggest a VWAP option pricing model based on modelling the underlying volume as the standard gamma process - an assumption which is backed by our empirical analysis of volume statistics of a few ASX-traded stocks. Our model allows us to obtain simple closed-form formulas for implied volatility adjustments (with no forward adjustments needed) for Black-Scholes-style pricing formulas. These formulas are excellent approximations to the VWAP gamma model exact results, obtained by Monte Carlo simulations. In addition we have demonstrated that some rather conventional approximations (see Eqs. (\ref{Ratios})--(\ref{Ratios2})), which our model can avoid, should be used with a great deal of care since they may lead to substantial pricing errors.

It is also reasonably clear how one could further improve on our results. One should aim to derive a more sophisticated model for the volume process which should have a richer parameter space to better fit the empirical stock volume data, than the simple gamma distribution utilised in this work and/or take into account the partial correlation between volume and stock processes. Another possible direction is to obtain upper/lower bound results based on the gamma distributed volume model.

\section*{Appendix A} \label{AppendixA}

We model the volumes $V_i$ traded within $[t_{i-1},t_{i}]$ time interval (where $t_i=i\Delta t$) by i.i.d.\! gamma variables $V_i\sim\Gamma(\alpha,\theta)$, independent of the underlying stock process $S_t$. Note that the parameter $\alpha$ depends linearly on the choice of $\Delta t$.

The vector of scaled increments $X_i=V_i/\sum_{j=1}^N V_j$ are given by
\[
\left(X_1,\ldots,X_N\right)=\left(\frac{V_1}{\sum_{j=1}^N V_j},\ldots,\frac{V_N}{\sum_{j=1}^N V_j}\right)\sim D(\alpha, \ldots, \alpha),
\]
where $D$ is a Dirichlet distribution. The following formulas for various expectations are well-known:
\begin{gather*}
\E(X_i)=\frac{1}{N}=\frac{\Delta t}{T},\quad \E(X_i^2)=\frac{\alpha + 1}{N(\alpha N+1)},\quad \Var(X_i)=\frac{N-1}{N^2(\alpha N+1)},\\
\E(X_i X_j)=\frac{\alpha}{N(\alpha N+1)},\quad \Cov(X_i,X_j)=\frac{-1}{N^2(\alpha N+1)},\quad i\neq j.
\end{gather*}

\subsection*{Discrete-time case}
In the discrete-time case, the VWAP is defined to be
\[
S := S^{\rm(VWAP)} =\frac{\sum_{i=1}^N S_iV_i}{\sum_{i=1}^N V_i}= \sum_{i=1}^N S_i X_i
\]
where $S_i=S_{i\Delta t}$.
The first two moments can then be computed using Fubini's Theorem, Fubini (1958): 
\begin{align*}
M_1&=\E S= \sum_{i=1}^N \E S_i \E X_i
=\E\left(\frac{1}{N}\sum_{i=1}^N S_i\right),\\
\E S^2 &=\sum_{i=1}^N \E S_i^2 \E X_i^2 + 2\sum_{i< j} \E (S_i S_j) \E (X_i X_j) \\
&= \frac{\alpha + 1}{N(\alpha N+1)} \sum_{i=1}^N \E S_i^2 + \frac{2\alpha}{N(\alpha N+1)} \sum_{i< j} \E (S_iS_j) \\
&= \E\left( \frac{1}{N}\sum_{i=1}^N S_i\right)^2 +  \frac{1}{\alpha N+1}\left( \E\left(\frac{1}{N}\sum_{i=1}^N S_i^2\right)-\E\left( \frac{1}{N}\sum_{i=1}^N S_i\right)^2\right),\\
M_2&=\E S^2-M_1^2\\
&= \Var \left( \frac{1}{N}\sum_{i=1}^N S_i\right) +  \frac{1}{\alpha N +1}\left( \E\left(\frac{1}{N}\sum_{i=1}^N S_i^2\right)-\E\left( \frac{1}{N}\sum_{i=1}^N S_i\right)^2\right).
\end{align*}

Note that instead of solving gamma process VWAP problem exactly as above, we can adopt the ``Stace-like'' approach as well, by utilising the approximations (\ref{Ratios})--(\ref{Ratios2}). In this case $M_1^{\rm (Stace)} = M_1$, but the final expression for $M_2$ changes into:
\begin{align*}
M_2^{\rm (Stace)} &= \Var \left( \frac{1}{N}\sum_{i=1}^N S_i\right) +  \frac{1}{\alpha N}\left( \E\left(\frac{1}{N}\sum_{i=1}^N S_i^2\right)-\left[\E\left( \frac{1}{N}\sum_{i=1}^N S_i\right)\right]^2 \right).
\end{align*}

Now let us recall that $S_t$ is a geometric Brownian motion with drift $r$ and volatility $\sigma$, all terms in the expressions for $M_1$, $M_2$ and $M_2^{\rm (Stace)}$ can all be explicitly calculated:
\begin{align*}
\E\left(\frac{1}{N}\sum_{i=1}^N S_i\right)&=\frac{S_0}{N}\sum_{i=1}^N e^{r i\Delta t},\\
\E\left(\frac{1}{N}\sum_{i=1}^N S_i^2\right)&=\frac{S_0^2}{N}\sum_{i=1}^N e^{(2r+\sigma^2) i\Delta t},\\
\E\left(\frac{1}{N}\sum_{i=1}^N S_i\right)^2&=\frac{S_0^2}{N^2}\left(\sum_{i=1}^N e^{(2r+\sigma^2) i\Delta t}+2\sum_{j=1}^N\sum_{i=1}^{j-1}e^{r(i+j)\Delta t+\sigma^2 i\Delta t}\right).
\end{align*}
This, in turn, allows the calculation of the corresponding implied volatilities by moment matching. For example, for equidistant spacing of averaging points, the following expressions can be obtained:
\begin{align*}
\sigma^{\rm (VWAP)}_{\rm ExactD} = \sigma \sqrt{\frac{(1+N) (3 + \alpha + 2 \alpha N)}{6 N (1 + \alpha N)}+ O(T/N)},
\end{align*}
or (if taken relative to $\sigma^{(AA)}_D$):
\begin{align*}
\frac{\sigma^{\rm (VWAP)}_{\rm ExactD}}{\sigma^{(AA)}_D} = \sqrt{\frac{N (3 + \alpha + 2 \alpha N)}{(1+2 N) (1 + \alpha N)}+ O(T/N)} ,
\end{align*}
and
\begin{align*}
\sigma^{\rm (VWAP)}_{\rm StaceD} = \sigma \sqrt{\frac{(1+N) (3 + \alpha + 2 \alpha N)}{6 \alpha N^2}+ O(T/N)},
\end{align*}
or (if taken relative to $\sigma^{(AA)}_D$):
\begin{align*}
\frac{\sigma^{\rm (VWAP)}_{\rm StaceD}}{\sigma^{(AA)}_D} = \sqrt{\frac{(3 + \alpha + 2 \alpha N)}{(\alpha + 2 \alpha N)}+ O(T/N)},
\end{align*}

\subsection*{Continuous-time case}
The continuous-time VWAP is the limit of the discrete-time case as $N\rightarrow \infty$ or $\Delta t\rightarrow 0$:
\[
\tilde{S}:=S^{\rm(VWAP)}=\frac{\int_0^T S_t\, dZ_t}{Z_T}=\lim_{N\rightarrow \infty} \sum_{i=1}^N S_{(i-1)\Delta t} X_i.
\]
where \emph{the aggregate volume traded} $Z_t$ is a gamma process $\Gamma(t;\tilde \alpha, \theta)$.
Here we use the parameter $\tilde \alpha$ which relates to the discrete-time case parameter via $\tilde \alpha = \alpha / \Delta t$ or $\tilde \alpha T= \alpha N$. Using arguments similar to the discrete-time case, as well as the dominated convergence theorem, the first two exact moments are:
\begin{align*}
M_1&=\E\left(\frac{1}{T}\int_0^T S_{t}\, dt\right),\\
\E \tilde{S}^2 &= \E\left(\frac{1}{T}\int_0^T S_t\, dt\right)^2 + \frac{1}{\tilde{\alpha} T+1}\left(\E\left(\frac{1}{T}\int_0^T S_t^2\, dt\right)- \E\left(\frac{1}{T}\int_0^T S_t\, dt\right)^2\right),\\
M_2 &=\Var\left(\frac{1}{T}\int_0^T S_t\, dt\right) + \frac{1}{\tilde{\alpha} T+1}\left(\E\left(\frac{1}{T}\int_0^T S_t^2\, dt\right)- \E\left(\frac{1}{T}\int_0^T S_t\, dt\right)^2\right).
\end{align*}
And again we can use \eqref{Ratios}--\eqref{Ratios2} to obtain a Stace-like approximation,
\[
M_2^{\rm (Stace)}=\Var\left(\frac{1}{T}\int_0^T S_t\, dt\right) + \frac{1}{\tilde{\alpha} T}\left(\E\left(\frac{1}{T}\int_0^T S_t^2\, dt\right)- \left[\E\left(\frac{1}{T}\int_0^T S_t\, dt\right)\right]^2\right).
\]

The terms involving the stock process $S_t$ are given by:
\begin{align*}
\E\left(\frac{1}{T}\int_0^T S_t\, dt\right)&=\frac{S_0}{T}\left(\frac{e^{r T}-1}{r}\right),\\
\E\left(\frac{1}{T}\int_0^T S_t\, dt\right)^2&=\frac{2S_0^2}{(r+\sigma^2)T^2}\left(\frac{e^{(2r+\sigma^2)T}-1}{2r+\sigma^2}-\frac{e^{r T}-1}{r}\right),\\
\E\left(\frac{1}{T}\int_0^T S_t^2\, dt\right) &= \frac{S_0^2}{T}\left(\frac{e^{(2r+\sigma^2)T}-1}{2r+\sigma^2}\right).
\end{align*}
Finally the corresponding implied volatilities for the continuous limit can also be calculated:
\begin{align*}
\frac{\sigma^{\rm (VWAP)}_{\rm Exact}}{\sigma^{(AA)}} = \sqrt{\frac{(3 + 2 \tilde{\alpha})}{(2 + 2 \tilde{\alpha})}+ \frac{(6 (1 + \tilde{\alpha}) r^2 + 3 (1 + \tilde{\alpha}) r \sigma^2 +
\tilde{\alpha} \sigma^4) T}{24 (1 + \tilde{\alpha})^2 \sigma^2} + O(T^2)} ,
\end{align*}
and
\begin{align*}
\frac{\sigma^{\rm (VWAP)}_{\rm Stace}}{\sigma^{(AA)}} = \sqrt{\frac{(3 + 2 \tilde{\alpha})}{2 \tilde{\alpha}}+ \frac{(2 \tilde{\alpha} r^2 + \tilde{\alpha} r \sigma^2 - 3 \sigma^4 -
\tilde{\alpha} \sigma^4) T}{8 \tilde{\alpha}^2 \sigma^2} + O(T^2)}.
\end{align*}

\subsection*{Acknowledgment}
The authors are grateful to A. Barber, A. Brace, T. Glass, V. Frishling, D. Maher, T. Ling, A. Novikov and W. Wright for fruitful discussions and useful suggestions.


\end{document}